\documentclass[nofootinbib]{revtex4}
\usepackage[english]{babel}
\usepackage{array,booktabs}   
\usepackage{array} 
\usepackage{amsmath} 
\usepackage{relsize}
\usepackage{wrapfig}
\usepackage{calc}
\usepackage{pdflscape}
\usepackage{color}
\usepackage{float}
\usepackage[font=small,labelfont=bf]{caption}
\usepackage{graphicx}
\usepackage{amsmath}
\usepackage[nodisplayskipstretch]{setspace}

\usepackage{setspace}
\usepackage{tabularx}

\usepackage{float}
\usepackage{color}
\usepackage{amsmath}
\usepackage{float}
\usepackage{calc}
\usepackage{pdflscape}
\usepackage{color}
\usepackage{float}
\usepackage{graphics}
\usepackage{epsfig}
\usepackage{epstopdf}
\usepackage{amssymb}
\usepackage[font=small,labelfont=bf]{caption}
\usepackage{graphicx}
\usepackage{soul}
\usepackage{epstopdf}
\usepackage[urlcolor=blue]{hyperref}
\begin{document}{\setlength\abovedisplayskip{4pt}}

\title{Constraining the Effective Mass of Majorana Neutrino with Sterile Neutrino Mass for Inverted Ordering Spectrum.}
\author{Jaydip Singh\footnote{E-mail: jaydip.singh@gmail.com}}

\affiliation{Department Of Physics, Lucknow University, $Lucknow-226007^{\ast,\dagger}$}

\begin{abstract} 
 
 Inspired by the experimental anomalies in neutrino physics and recent oscillation data from short baseline and another neutrino experiment, the realization of one 
 extra neutrino flavor seem to be favoring. This extra flavor may change the observable, $|m_{\beta\beta}|$ of currently data taking and next-generation $(\beta\beta)_{0\nu}$-decay 
 experiments aim to probe and possibly look the Inverted Ordering region($|m_{\beta\beta}| \simeq 10^{-2}$eV) of parameter space. This observation would allow 
 establishing physics beyond the standard model and phenomena like lepton number violation and Majorana nature of neutrino. The range of this observable ($|m_{\beta\beta}|$) 
 is not very well defined for both the ordering of mass spectrum(Normal Ordering and Inverted Ordering). Several attempts have been made for defining exactly the range 
 for three active neutrino states. For contrasting this range, I have worked with an extra mass states, $\nu_{4}$ and its effect on the observable with various 
 combination of CP violation Majorana phases by taking into account the updated data on the neutrino oscillation parameters for IO case. Based on the monte carlo 
 technique, a parameter region is obtained using the fourth Majorana-Dirac phase of sterile parameters that lead to an effective mass below 0.01 eV or .05 eV for 
 inverted mass ordering case which is planned to be observed in the near future experiment.  

\end{abstract}
\maketitle

\section {Introduction}	
\setlength{\baselineskip}{13pt}

 In future, the positive observation of neutrinoless double beta decay process would be a clear evidence of lepton number violation and confirmation of the Majorana 
 nature of neutrinos\cite{ref1}. In addition to the phenomena discussed earlier, it can also address some of the yet unresolved issues in neutrino physics and physics 
 beyond the standard model, such as the origin of tiny neutrino masses, the absolute neutrino mass scale and the mass ordering, see Refs.\cite{ref2,ref21,ref3,ref4} and 
 also dark matter physics. In general, three active neutrinos are considered in the standard neutrino oscillation picture with mass-squared differences of order 10$^{-4}$ 
 and 10$^{-3}$ eV$^{2}$\cite{ref09}. From tritium beta decay experiments and cosmological observations, the absolute mass scale of neutrino is also constrained below 
 around 1 eV. Considering the three neutrino formalism we have various successful achievements of atmospheric, solar, accelerator and reactor neutrino experiments but 
 there are experimental anomalies that cannot be explained within the standard three-neutrino framework. Mainly the issue of LSND\cite{ref001} and MiniBooNE\cite{ref002} 
 results that are not yet explained clearly is frequently interpreted as a hint towards the existence of one or two sterile neutrino states with masses at the eV 
 scale\cite{ref001,ref002}. Such neutrinos called "sterile" because it cannot participate in the weak interactions due to collider constraints and points towards the 
 non-standard neutrino physics. The presence of sterile neutrinos would significantly change the observables in neutrino experiments, specifically the oscillation 
 probabilities in short-baseline experiments and the effective mass in neutrino-less double beta decay\cite{ref003,ref003a,ref004}. Recent new data from reactor 
 experiment, IceCube, MINOS+ and global $\nu_{e}$ disappearance channel supports the sterile neutrino explanation of the reactor anomalies and anomalies of other 
 short-baseline neutrino oscillation experiments\cite{ref005}.
 
In the oscillation experiment, total lepton number is conserved while in the neutrinoless double beta decay experiment total lepton number changes by two units. So, 
some of the fundamental question like lepton number violation and Majorana nature of neutrino cannot be answered through the neutrino oscillation experiments but can 
be answered through the Neutrino-less double beta decay (0$\nu\beta\beta$) experiment. Therefore the most promising process which can unambiguous the Majorana nature 
of neutrinos is neutrinoless double beta decay and the search for this phenomena have a long history\cite{ref01}. Current running experiments like KamLAND-Zen, CUORE, 
 CUORE-0, Cuoricino, and GERDA-II are trying to find out the lower bounds on the $T_{1/2}^{0\nu}$ of this decay for several nucleus samples. Recent lower bound obtained 
 by the KamLAND-Zen collaboration for sample $^{136}Xe$(xenon-136) is, $T_{1/2}^{0\nu} > 1.07 \times 10^{26}$ yr\cite{ref04}, GERDA-II collaboration obtained the 
 lower bound for sample $^{76}Ge$(germanium-76) is $T_{1/2}^{0\nu} > 8.0 \times 10^{25}$ yr\cite{ref03} and CUORE, CUORE-0 and Cuoricino experiments collectively 
 obtained the lower bound for sample $^{130}Te$(tellurium-130) is $T_{1/2}^{0\nu} > 1.5 \times 10^{25}$ yr\cite{ref03}.

 The $(\beta\beta)_{0\nu}$ - decay rate is proportional to the effective Majorana mass $|m_{\beta\beta}|$ in the three Majorana neutrino picture. The range of this 
 effective Majorana mass is not very well understood but based on the present neutrino oscillation data, it is bounded from below $|m_{\beta\beta}|_{IO} > 1.4 
 \times 10^{-2}$ eV in the case of three neutrino mass spectrum with Inverted Ordering\cite{ref05}. While for the Normal Ordering configuration of the mass spectrum 
 the lower bound is $|m_{\beta\beta}|_{NO} << 10^{-3}$ eV\cite{ref06} and it can be extremely small depending on the values of the Dirac, Majorana phases and of the 
 smallest neutrino mass. A condition is established using the recent global data on the neutrino oscillation parameters with NO(Normal Oredering) spectrum\cite{ref010}, 
 which suggest $|m_{\beta\beta}|$ must exceed $10^{-3}(5\times10^{-3})$ eV. Neutrino double beta decay experiments are trying to cover the range of $|m_{\beta\beta}|$ from 
 the top and the current reachable upper limit reported by KamLAND-Zen collaboration is $|m_{\beta\beta}| < (0.061-0.165)eV$\cite{ref04}. This upper limits is obtained by 
 the KamLAND-Zen collaboration using the lower limit on the half-life of the xenon-136 sample and they have considered the uncertainties in the NMEs(Nuclear Matrix 
 Elements) in their analysis of the relevant process.
    
New-generation experiments plan to look the Inverted Ordering region of parameter space and possibly work for the $|m_{\beta\beta}| \sim 10^{-2}$ eV energy range. 
 Various running experiments are considered\cite{ref011,ref012} for upgradation and new experiments are also proposed to achive this goal. Some of those experiments 
 are as fallows, MAJORANA, LEGEND$(^{76}Ge)$, CANDLES$(^{48}Ca)$, AMoRE, PandaX-III, SuperNEMO and DCBA$(^{82}Se, ^{150}Nd)$, ZICOS $(^{96}Zr)$, MOON $(^{100}Mo)$, 
 COBRA $(^{116}Cd, ^{130}Te)$, SNO+$(^{130}Te)$, NEXT and nEXO$(^{136}xE)$. If these planned experiments didn't find positive responce in this energy range($10^{-2} eV$), 
 then next generation experiment will be very intersting for sterile neutrino which will correspond $|m_{\beta\beta}|\sim 10^{-3}$ eV or more below in the 
 $(\beta\beta)_{o\nu}$-decay experiment.  
 
 The introduction of a sterile neutrino at the eV mass scale can change the prediction for the possible range of $|m_{\beta\beta}|$ values in the neutrinoless 
 double-beta decay\cite{ref003,ref013,ref014,ref015,ref016,ref017}. In the present article, I have determined a sterile parameter region using monte carlo technique 
 under which the effective Majorana mass is pushed below a certain value (.01 eV or .05 eV) in the case of 1+3 neutrino mixing with Inverted Ordering mass spectrum. 
 For completeness Normal Ordering scenario of the spectrum with three active and one sterile neutrino is also discussed with the recent global data. Considering the 
 latest global data and various possible CP violation Majorana phases possible ranges of observables in $(\beta\beta)_{0\nu}$-decay, tritium beta decay experiments 
 and cosmological observations are also discussed.

\section{THE EFFECTIVE MASS IN 3 and 1+3 NEUTRINO STATES} 
  In this section, I will discuss the general formalism of neutrino parameters and evaluate the contributions to the effective mass relevant for neutrino-less 
  double beta decay. First I outline the formalism of three active neutrino mixing and then mixing in the presence of one extra sterile states (1+3), as discussed 
  in\cite{ref003}. I have worked with the 1+3 scenario when the sterile neutrino is heavier than the active neutrino with Normal Ordering and Inverted Ordering of the 
  spectrum, other possible scenarios (3+1, 3+2/2+3 or 1+3+1) are also available in the literature but not discussed here, more detail information can be found in 
  reference\cite{ref003}.  
  
   \subsection{THREE ACTIVE NEUTRINO MIXING }
 For three acive neutrino states configuration the effective Majorana mass $|m_{\beta\beta}|$ is defined as : 
  \begin{equation}
   |m_{\beta\beta}| = |\sum_{k=1}^{3}U_{ek}^{2} \mu_{k}|
  \end{equation}
  
  with the partial contribution of the massive Majorana neutrino $\nu_{k}$ with mass $\mu_{k}$ and U being the leptonic mixing matrix and known as Pontecorvo-Maki-
  Nakagawa-Sakata(PMNS) matrix, which exactly defined the mixing of the electron neutrino with the three massive neutrinos. The first row of this mixing matrix is 
  the one relevent for $(\beta\beta)_{0\nu}$-decay, and In standard parametrization it is defined\cite{ref06} as,
  \begin{equation}
   U_{ek} =(c_{13}c_{12}, c_{13}s_{12}e^{i\phi_{21}/2}, s_{13}e^{-i\delta}e^{i\phi_{31}/2})
  \end{equation}

   where $c_{kl}$ $\equiv$ $cos\theta_{kl}$ and $s_{kl}$ $\equiv$ $sin\theta_{kl}$, where $\theta_{kl}\in [0,\pi/2]$ are the mixing angles, and $\delta$ and the 
   $\phi_{kl}$ are the Dirac and Majorana phases\cite{ref04a}, respectively and these are the coplex phases and the possible range of these parameters are $[0,2\pi]$. 
   The values of these phases is not known and all possible values of $|m_{\beta\beta}|$ must take into account all the possible range of these phases. The results 
   for the neutrino squared-mass differences are expressed in terms of the solar and atmospheric squared mass differences, which are defined as 
     \begin{equation}
      \Delta m_{sol}^{2} = \Delta \mu_{21}^{2}
     \end{equation}
     \begin{equation}
      \Delta m_{atm}^{2} = \frac{1}{2}|\Delta \mu_{31}^{2} + \Delta \mu_{32}^{2}|
     \end{equation}
     where $\Delta \mu_{jk}^{2} = \mu_{j}^{2} - \mu_{k}^{2}$ . Given this assignment of the squared mass differences, it is currently unknown if the ordering 
     of the neutrino masses is normal, i.e. $\mu_{1} < \mu_{2} < \mu_{3}$ or inverted, i.e.  $\mu_{3} < \mu_{1} < \mu_{2}$. I also defined $m_{low}$ $\equiv$ 
     $\mu_{1}(\mu_{3})$ in the NO(IO). In terms of the lightest neutrino mass, CPV phases, neutrino mixing angles, and neutrino mass-squared differences, the 
     effective Majorana mass reads:
    \begin{equation}
     |m_{\beta\beta}|_{NO} = |m_{low}c_{12}^{2}c_{13}^{2}+\sqrt{\Delta m_{sol}^{2}+m_{low}^{2}}s_{12}^{2}c_{13}^{2}e^{i\phi_{21}}+\sqrt{\Delta m_{atm}^{2} +  m_{low}^{2}} s_{13}^{2} e^{i \phi _{31}^{'}}|
    \end{equation}
    
    \begin{equation}
     |m_{\beta\beta}|_{IO} = |\sqrt{\Delta m_{atm}^{2} -  \Delta m_{sol}^{2} + m_{low}^{2}}c_{12}^{2}c_{13}^{2} + \sqrt{\Delta m_{atm}^{2}+m_{low}^{2}}s_{12}^{2}c_{13}^{2}e^{i\phi_{21}} 
      + m_{low} s_{13}^{2} e^{i \phi _{31}^{'}}|,
    \end{equation}
    
     where we have defined $\phi_{31}^{'}$ $\equiv \phi_{31} - 2\delta $. Effective mass for both the case, $|m_{\beta\beta}|_{NO}$ and $|m_{\beta\beta}|_{IO}$, can be 
     written in the form
     \begin{equation}
      |m_{\beta\beta}| = |\tilde{\mu_{1}}+ \tilde{\mu_{2}}e^{i\phi_{21}}+ \tilde{\mu_{3}}e^{i\phi_{31}^{'}}|,
     \end{equation}
     
     It is then clear that the effective Majorana mass is the length of the vector sum of three vectors in the complex plane for three neutrino mixing case. And, their 
     relative orientations are determined by the CPV phase factor $\phi_{21}$ and $\phi_{31}^{'}$ that can push the range of $|m_{\beta\beta}|$ significantly up and down
     with Normal or Inverted ordering of the neutrino mass spectrum. Table I and II show the best fit, 1$\sigma$, 2$\sigma$ and 3$\sigma$ ranges of the three neutrino oscillation 
     parameters used for this analysis with NO and IO of the mass spectrum obtained from the recently updated global data\cite{ref09}.  

     Generally the observation of the absolute values of neutrino masses is done through the measurements of the effective electron neutrino mass in the $\beta$-experiment 
     and it is expressed as
     \begin{equation}
      m_{\beta} = \sqrt{|U_{e1}|^{2}\mu_{1}^{2} + |U_{e2}|^{2}\mu_{2}^{2} + |U_{e3}|^{2}\mu_{3}^{2}}
     \end{equation}
     
     and the sum of the neutrino masses in cosmological experiments, which is defined as : 
     \begin{equation}
      \sum = \mu_{1} + \mu_{2} + \mu_{3}
      \end{equation}

     Therefore, it is helpful to estimate the allowed regions in the $m_{\beta} - |m_{\beta\beta}|$ and $\sum - |m_{\beta\beta}|$ planes, as shown in the Figure 3.

\subsection{FOUR NEUTRINO MIXING CASE}

In this section I consider the case of 1+3 mixing in which there is a new massive neutrino $\nu_{4}$ at the eV scale which is mainly sterile. By accommodating one 
sterile neutrino in three neutrino mixing mechanism, the effective Majorana mass in 0$\nu\beta\beta$ is expressed as \cite{ref003} :

    \begin{equation}
     |m_{\beta\beta}|_{4\nu} = |c_{12}^2 c_{13}^2 c_{14}^2 \mu_{1}+s_{12}^2 c_{13}^2 c_{14}^2 \mu_{2}e^{i\phi_{21}}+s_{13}^2 c_{14}^2 \mu_{3}e^{i\phi_{31}^{'}}+ s_{14}^2 \mu_{4}e^{i\phi_{41}}|
    \end{equation}
    and this expression can be written as 
    \begin{equation}
     |m_{\beta\beta}| = |\tilde{\mu_{1}} + \tilde{\mu_{2}}e^{i\phi_{21}} + \tilde{\mu_{3}}e^{i\phi_{31}^{'}} + \tilde{\mu_{4}}e^{i\phi_{41}}|
    \end{equation}
    . This extra flavor $\nu_{4}$ contribution comes with an extra complex phase factor $\phi_{41}$ that must be varied between 0 to 2$\pi$  for full physics analysis 
    as $\phi_{21}$ and $\phi_{31}$ to estimate the all possible value of $|m_{\beta\beta}|$. However, for both the ordering(NO and IO), $\tilde{\mu_{4}}$ remains the 
    same and defined as : 
    \begin{equation}
     \tilde{\mu_{4}} \simeq \sqrt{m_{low}^{2} + \Delta m_{sbl}^{2}}s_{14}^{2}, 
    \end{equation}
    
    here one can neglect the contributions of $\Delta m_{sol}^{2}$ and $\Delta m_{atm}^{2}$, which are much smaller then $\Delta m_{sbl}^{2}$. If three active neutrino 
    is lighter than the sterile neutrino ones, the expression for $|m_{\beta\beta}|$ can be experssed as:
     \begin{equation}
       |m_{\beta\beta}|_{(1+3)\nu}  \simeq |c_{14}^{2}<m_{\beta\beta}>_{3\nu} + s_{14}^{2} \sqrt{\delta \mu_{41}^{2}}e^{i\phi_{41}}|
      \end{equation}
      where $<m_{\beta\beta}>_{3\nu}$ is the standard expression for three active neutrinos as defined by eq. (5) and (6). Table III shows the best-fit, 2$\sigma$ and 
      3$\sigma$ range of the oscillation parameter for fourth sterile states used in this work for this analysis. Right panel of Figure 1 and 2, shows the allowed range 
      of $|m_{\beta\beta}|_{1+3\nu}$ as a function of the lightest mass for NO and IO case respectively, using data from \cite{ref005} for 1+3 case and \cite{ref09} for 
      3 active neutrino state.

\section{NORMAL ORDERING CONFIGURATION}
    The aim of this work is to constrain the $|m_{\beta\beta}|$ range for IO spectrum only but for completeness I will start my analysis with the normal ordering of 
    the mass spectrum, discussed in this section, then in the following sections I will discuss the inverted ordering of mass spectrum in detail. The length of the 
    vector, that determines $|m_{\beta\beta}|$ with the phase factor can be obtained from the equation (11) and (12), and the vectors as a function of lightest neutrino 
    mass with NO are $\tilde{\mu_{1}} = m_{low} c_{12}^{2} c_{13}^{2}c_{14}^{2}$, $\tilde{\mu_{2}} = \sqrt{\Delta m_{sol}^{2} + m_{low}^{2}}s_{12}^{2}c_{13}^{2}c_{14}^{2}$, 
    $\tilde{\mu_{3}} = \sqrt{\Delta m_{atm}^{2} + m_{low}^{2}}s_{13}^{2}c_{14}^{2}$ and $\tilde{\mu_{4}} \simeq \sqrt{m_{low}^{2} + \Delta m_{sbl}^{2}}s_{14}^{2}$. The 
    lengths of the three partial mass contribution to $|m_{\beta\beta}|$ in Eq. (6) are shown in the left panel of Figure 1 as functions of lightest neutrino mass, 
    $m_{low}$ for 3$\sigma$ variations of oscillation parameters as provided in the table (I). One can see that contribution of $\mu_{4}$ is dominant for 
    $m_{low}$ $\leq$ .01 eV while $\mu_{2}$ is dominant for $m_{low}$ $\leq$ .001 eV and cannot be cancelled by the smaller contribution of $\mu_{1}$ and $\mu_{3}$ for 
    any values of the relative phase differences as shown in the right panel of Figure 1. In the other region cancellation are possible mainly between the values 
    .001 - .1 eV. Result for effective Majorana mass, $|m_{\beta\beta}|$ as a function of $m_{low}$ is shown in the right panel of Figure 1 with three active neutrino 
    state and one extra flavour state. The minimum and maximum values of the allowed bands for $|m_{\beta\beta}|$ is obtained from the relative phase and the 
    coefficients.

\begin{figure}[h!]
\centering
\includegraphics[scale=.5]{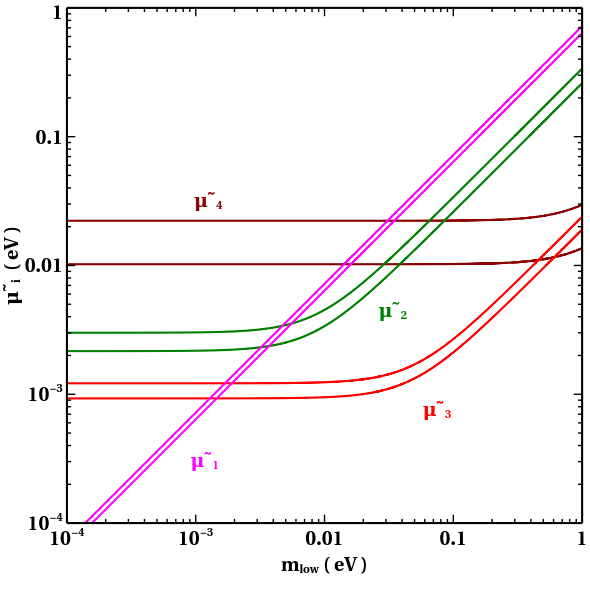}
\includegraphics[scale=.5]{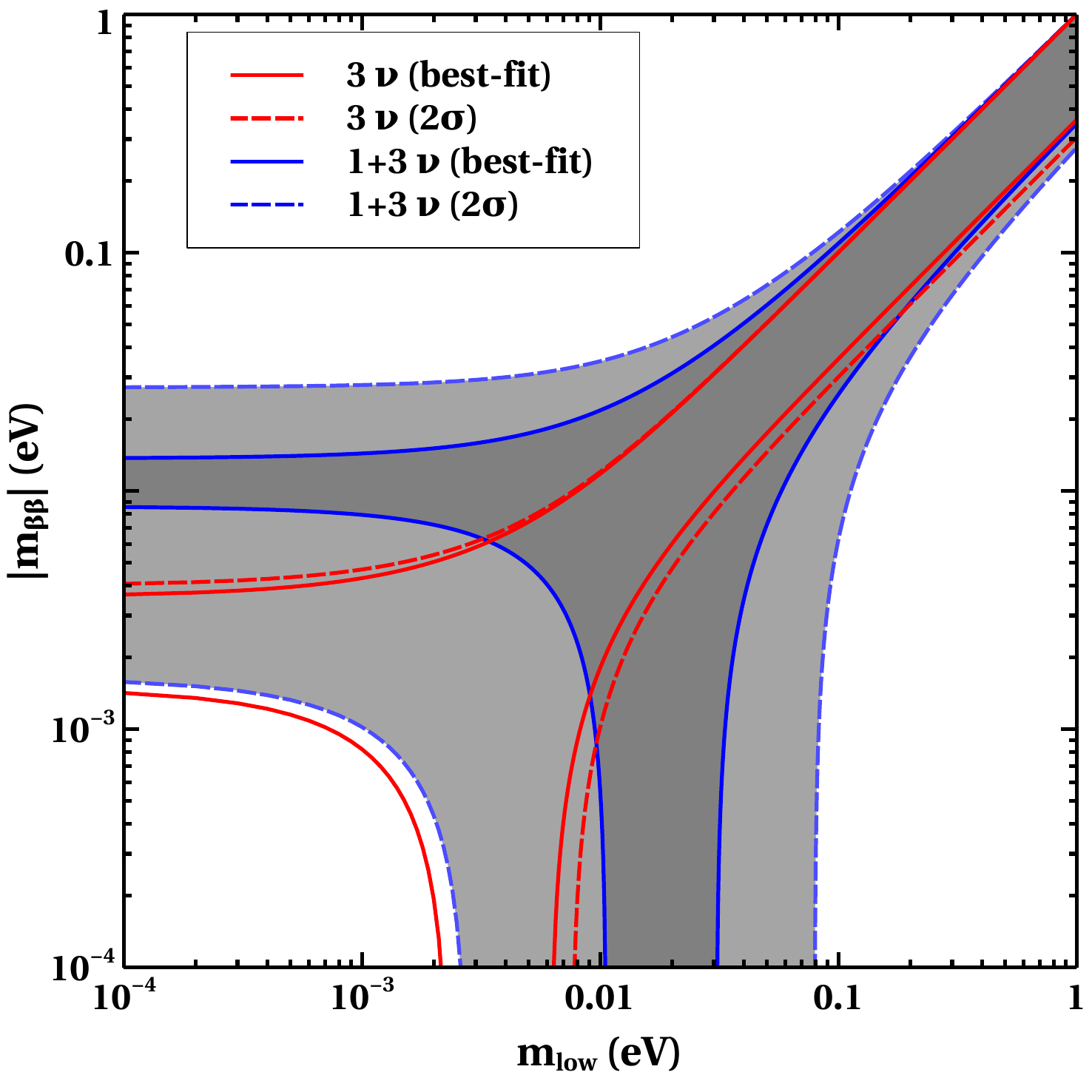}
\caption{ Left panel shows the four partial mass contribution to $|m_{\beta\beta}|$ in Eq. (12) as function of the lightest mass $m_{low}$ for $3\sigma$ allowed 
intervals in the case of 1+3$\nu$ mixing with Normal Ordering. Right panel shows the allowed ranges in the $|m_{\beta\beta}|$ - $m_{low}$ parameter space, both 
in the standard three neutrino picture(unfilled band space) and with one sterile neutrino (filled band space) for the 1+3 case as defined by eq. 14.}
\end{figure}

\section{INVERTED ORDERING CONFIGURATION}
 For 1+3 neutrino states with Inverted Ordering, the length of the vectors, that determines $|m_{\beta\beta}|$ with the phase factor can be obtained from equation 
 (6), (11) and (12). These vectors as a function of lightest neutrino mass $m_{low}$ are $\tilde{\mu_{1}} = \sqrt{|\Delta m_{atm}^{2}| - \Delta m_{sol}^{2} + m_{low}^{2}} c_{12}^{2} c_{13}^{2} c_{14}^{2}$, 
 $\tilde{\mu_{2}} = \sqrt{|\Delta m_{atm}^{2}| + m_{low}^{2}}s_{12}^{2}c_{13}^{2}  c_{13}^{2}$, $\tilde{\mu_{3}} = m_{low}s_{13}^{2} c_{14}^{2}$ and $\tilde{\mu_{4}} 
 \simeq \sqrt{m_{low}^{2} + \Delta m_{sbl}^{2}}s_{14}^{2}$. The lengths are shown in the left panel of Figure 2 as functions of $m_{low}$ for 3$\sigma$ variations of 
 oscillation parameters as given in the Table II and III. Right panel shows the value of effective Majorana mass $|m_{\beta\beta}|$ as a function of lightest neutrino mass 
 in the 3 and 1+3 neutrino case as defined by equation (6) and (13) respectively. One can see that $\mu_{1}$ is always dominant to, because $\theta_{13}$ is smaller than 
 $\pi/4$ and $|U_{e1}|>|U_{e2}|>|U_{e3}|$. Therefore, there can not be a complete cancellation of three mass contribution to  $|m_{\beta\beta}|$ for three active 
 neutrinos with Inverted Ordering, as shown in the right panel of Figure 3. So one can obtain the lower and upper bounds of $|m_{\beta\beta}|$ for three active neutrino 
 states with Inverted Ordering. Here, maximum and minimum values of the allowed bands for $|m_{\beta\beta}|$ is obtained from the relative phase factor and the 
 coefficients of the oscillations parameters as given in Table II. From the right panel of Figure 2, one can obtain the lower and upper bounds of the $|m_{\beta\beta}|$ 
 in the case of three active neutrino with Inverted Ordering mass spectrum,  
    \begin{equation}
     .02 eV \leq |m_{\beta\beta}| \leq .05  (best-fit) eV.
     \end{equation}
  The next generation and current data taking neutrinoless double-beta decay experiments will try to explore the range of $|m_{\beta\beta}|$ between the limits in Eqs. 
  (15) and testing the Majorana nature of neutrinos in the case of an Inverted Ordering.

 \begin{figure}[h!]
\centering
\includegraphics[scale=.5]{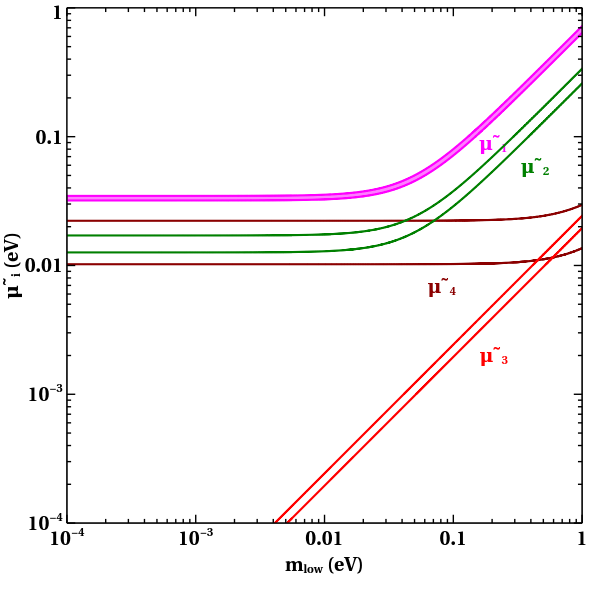}
\includegraphics[scale=.5]{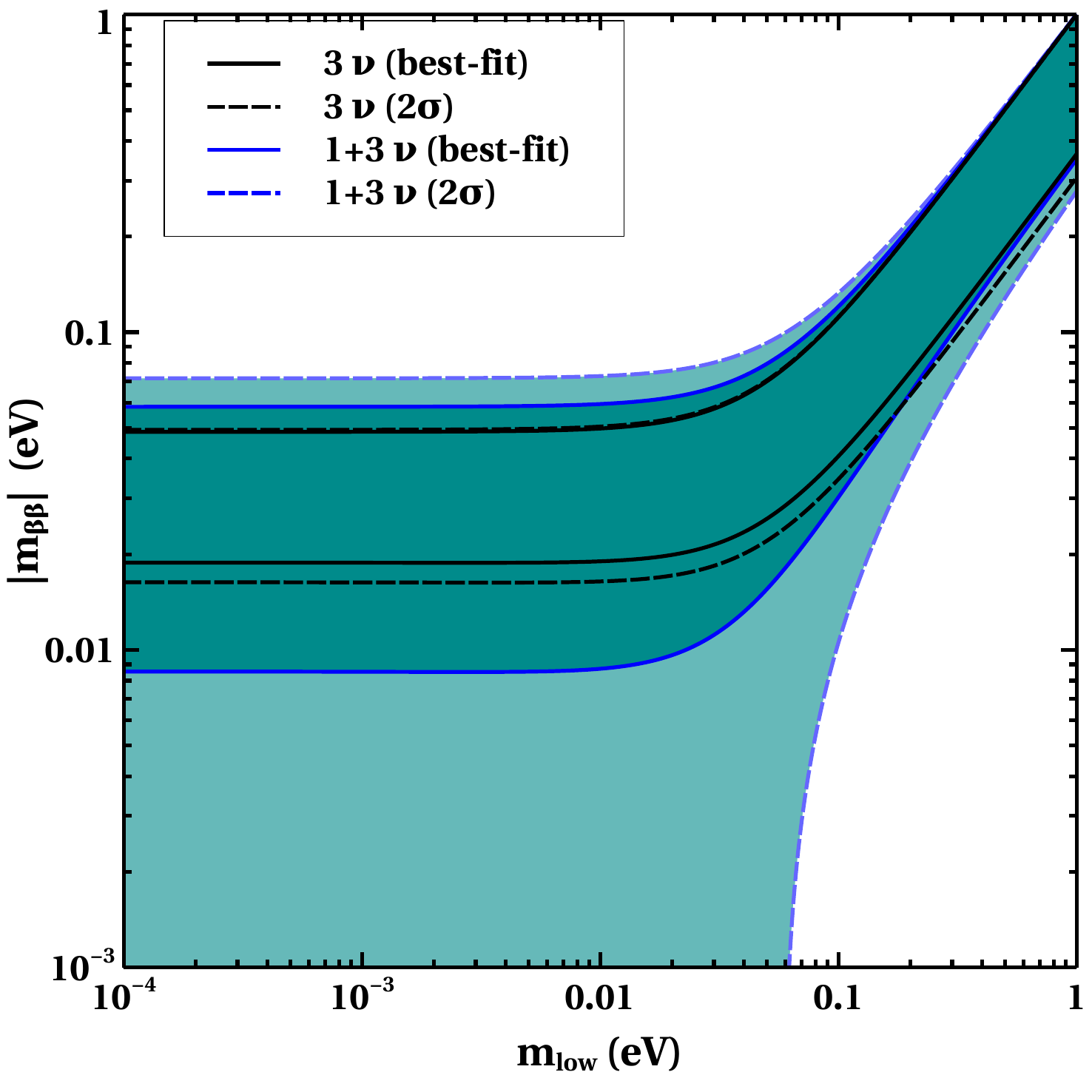}
\caption{ Left pannel shows the four partial mass contribution to $|m_{\beta\beta}|$ in Eq. (12) as function of the lightes mass $m_{low}$ for $3\sigma$ allowed 
intervals in the case of 1+3$\nu$ mixing with Inverted Ordering. Right pannel shows the allowed ranges in the $|m_{\beta\beta}|$ - $m_{low}$ parameter space, both 
in the standard three neutrino picture(unfilled band space) and with one sterile neutrino (filled band space) for the 1+3 case as defined by eq. (14).}
\end{figure} 

 While for the sterile neutrino case there can be a total cancellation between the partial contribution $\mu_{4}$ and $\mu_{2}$ for $m_{low} \leq$ .06 eV as shown in 
 the right panel of Figure 2. Comparing 3$\nu$ and 1+3$\nu$ mixing as shown in the right panel of Figure 2, One can see that the predictions of effective Majorana mass 
 $|m_{\beta\beta}|$ are completely different in the 3$\nu$ and 1+3$\nu$ cases if there is an Inverted Ordering\cite{ref003,ref013,ref014,ref015,ref016,ref017,ref018,ref019}. 
 This deifference in the effective Majorana mass $|m_{\beta\beta}|_{1+3\nu}$ comes mainly from the phase factors, $\phi_{21}$, $\phi_{31}^{'}$ and $\phi_{41}$. 
 Various possible range of $|m_{\beta\beta}|$ for the particular choice of the phase cases are shown in the Appendix.  

  \begin{figure}[h!]
\centering
\includegraphics[scale=.5]{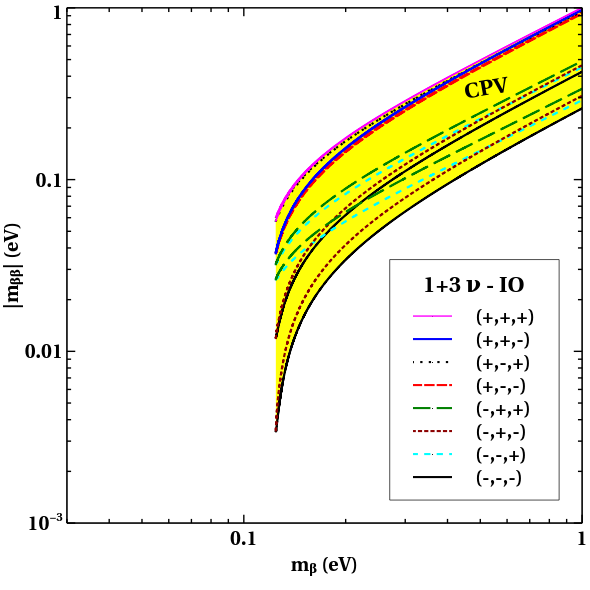}
\includegraphics[scale=.5]{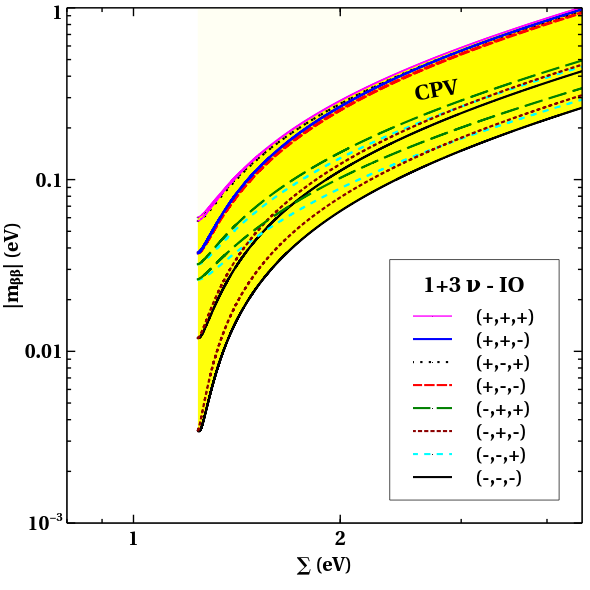}
\caption{ Left and right pannel respectively shows the value of effective Majorana mass $|m_{\beta\beta}|$ as a function of effective electron neutrino mass $m_{\beta}$ 
and sum of the neutrino masses $\sum$ in the case of 1+3 neutrino mixing with Inverted Ordering. The signs combination as shown in the legend imply the signs of 
$e^{i\phi_{21}}, e^{i\phi_{31'}}, e^{i\phi_{41}} = \pm$ 1 for the eight possible cases in which CP is conserved.}
\end{figure}
  
  Since $|m_{\beta\beta}|$ strongly depend on the phases $\phi_{21}$, $\phi_{31}^{'}$ and $\phi_{41}$ and, this phase combination determine the maximum and minimum of 
  $|m_{\beta\beta}|$ with coefficients of oscillation parameters. These phase combinations can significantly change the range of the $|m_{\beta\beta}|$, region of 
  $|m_{\beta\beta}|- m_{\beta}$ plane and region of $|m_{\beta\beta}|- \sum$ plane as shown in the Figure 2(right panel and 5 ) and 3(right and left panel) respectively. 
  Here I consider the phase combination which give the maximum and minimum possible range of $|m_{\beta\beta}|$ as shown in the right panel of Figure 2 and its 
  dependency on the fourth sterile phase parameter $(\phi_{41})$ only for estimating the lower bound. For understanding these phase dependency on $|m_{\beta\beta}|$, a 
  monte carlo code is developed, which is based on root package\cite{ref019aa}. In the code all parameters are varied in the full range to get the maxima and minima of 
  $|m_{\beta\beta}|$. Than to find the particular parameter region, in my case  ($\phi_{41}$ vs $m_{low}$), as shown in figure 4, that range causes the 
  $|m_{\beta\beta}|$ below a certain values is estimated. I have varied the full range of $\phi_{41}$, i.e. $(0,2\pi)$ and lightest neutrino mass in the range of 
  1.00 - .0001 eV and then find a region between $\phi_{41}$ and $m_{low}$ that can push the effective mass below the certain values (.01 eV or .05 eV) shown in 
  the Figure 4. If these parameter lies within the red dotted region, the effective Majorana mass will be less then .01 eV, similarly if these parameter lies within 
  blue dotted region, $|m_{\beta\beta}|$ will be $\leq$ .05 as shown in Figure 4. If the scenario 1+3$\nu$ mixing with Inverted Ordering of mass spectrum exist and 
  $|m_{\beta\beta}|$ goes below .01 eV or .05 eV and observed in the near future then sterile parameter must lies within the region shown in Figure 4.
  
  For completeness left and right panel of Figure 3 shows the correlation between $|m_{\beta\beta}|$ and the measurable quantities $m_{\beta}$ and $\sum$ for all the 
  possible combination of phases, that will be observed in beta decay experiment\cite{ref019a,ref019b,ref019c}, cosmological and astrophysical data\cite{ref019d,ref019e} 
  respectively.

\section{Result and Discussion}

\begin{figure}[h!]
\centering
\includegraphics[scale=.5]{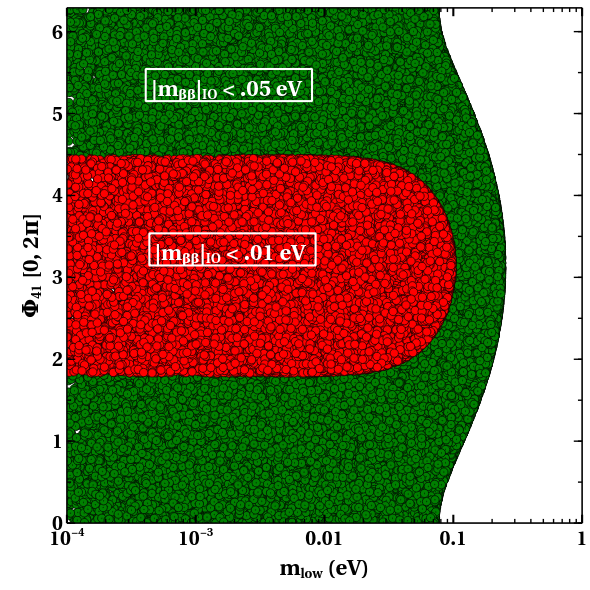}
\caption{ Regions in the $(m_{low}, \phi_{21})$ plane where two different conditions on $|m_{\beta\beta}|_{IO}$ apply. In the green region $|m_{\beta\beta}|_{IO}$ 
satisfies $|m_{\beta\beta}|_{IO} < .05 eV $ and red region satisfies $|m_{\beta\beta}|_{IO} < .01 eV$, for all values of $\theta_{kl}$, $\delta \mu_{jk}^{2}$ and 
$\phi_{31}^{'}$ from the corresponding $2\sigma$ intervals of oscillation parameter.}
\end{figure} 

Current data taking and next-generation $(\beta\beta)_{0\nu}$ - decay experiments aiming to look and possibly cover the Inverted Ordering range $|m_{\beta\beta}| 
\simeq 10^{-2}$ eV as obtained for three neutrino case. For contrasting this range in four neutrino case by considering recent global-fit data on the neutrino 
mixing angles and the neutrino mass-squared differences, I have determined the parameter space under which the effective Majorana mass in the Inverted Ordering 
case $|m_{\beta\beta}|_{IO}$ goes down to .05 eV and .01 eV as shown in Figure 4. For obtaining this parameter space I have considered the full range$(0,2\pi)$ of 
fourth Dirac-Majorana phase $\phi_{41}$ while keeping other phase fixed and varying the lightest neutrino mass in the range of 1.00 eV  -  1$\times 10^{-4}$ eV.

Finally one can conclude now by saying that if we obtained the $|m_{\beta\beta}|$ below the range discussed then it will be very interesting for sterile neutrino 
physics. And the parameter range obtained in this analysis will be useful for the development of models and experiments in $10^{-3}$ eV range to explore the sterile 
neutrino physics. Several oscillation experiments are planned mainly to investigate the LSND/MiniBooNE anomaly, particularly the Short-Baseline Neutrino program at 
DUNE (Deep Underground Neutrino Experiment) at Fermilab, which will use three detectors: Short-Baseline Near Detector\cite{ref022}, MicroBooNE\cite{ref020}, and ICARUS\cite{ref021}. 

\section{Acknowledgment} 
This work is partially supported by Department of Physics, Lucknow University, Lucknow, India. Financially it is supported by Government of India, DST Project no-SR/MF/PS02/2013, 
 Department of Physics, Lucknow Lucknow University. I thank Prof. Werner Rodejohann for useful discussion and suggestions and Dr. Jyotsna Singh for providing various 
 support in the department.

\appendix
\section{IO case with Other Phase }
         
 \begin{figure}[h!]
\centering
\includegraphics[scale=.5]{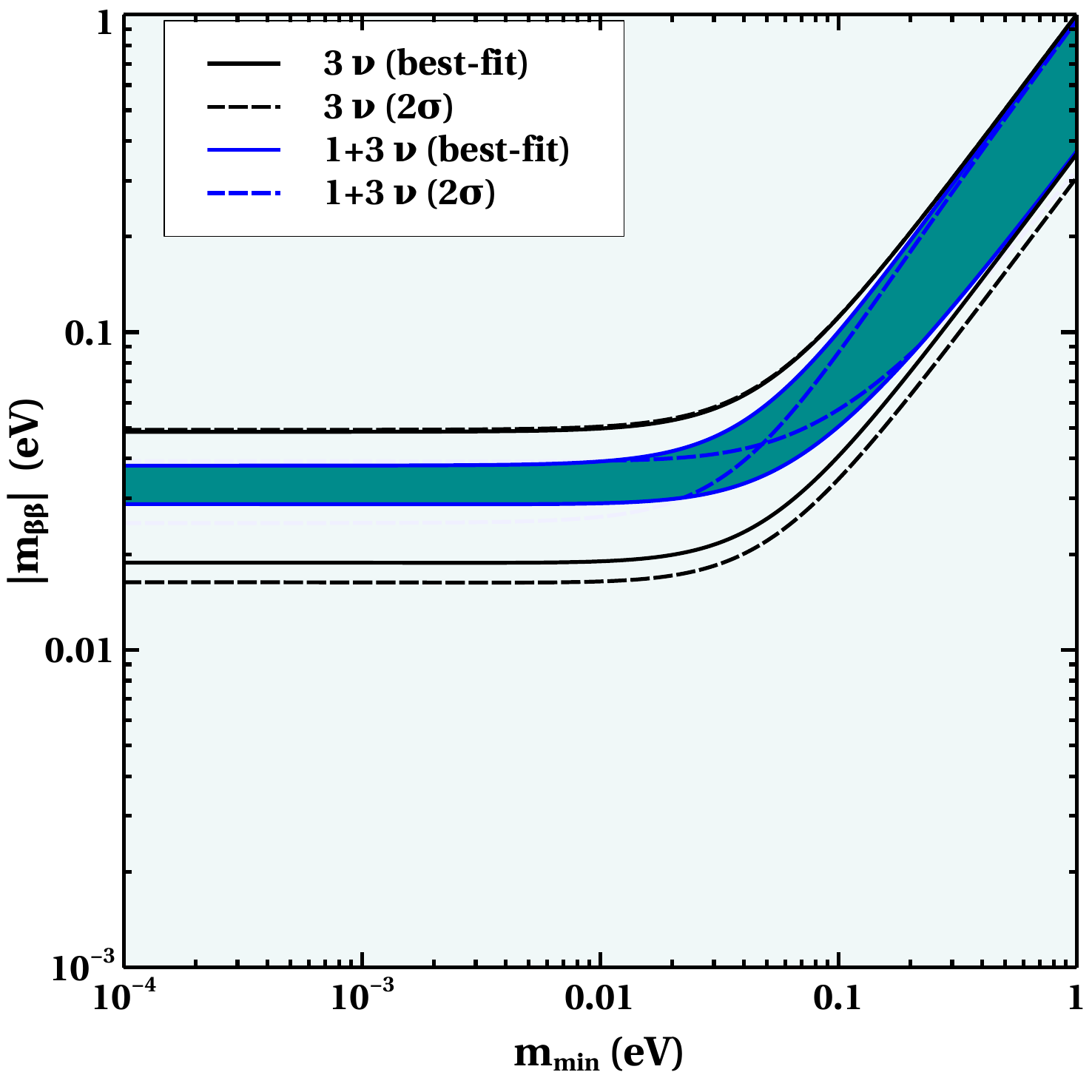}
\includegraphics[scale=.5]{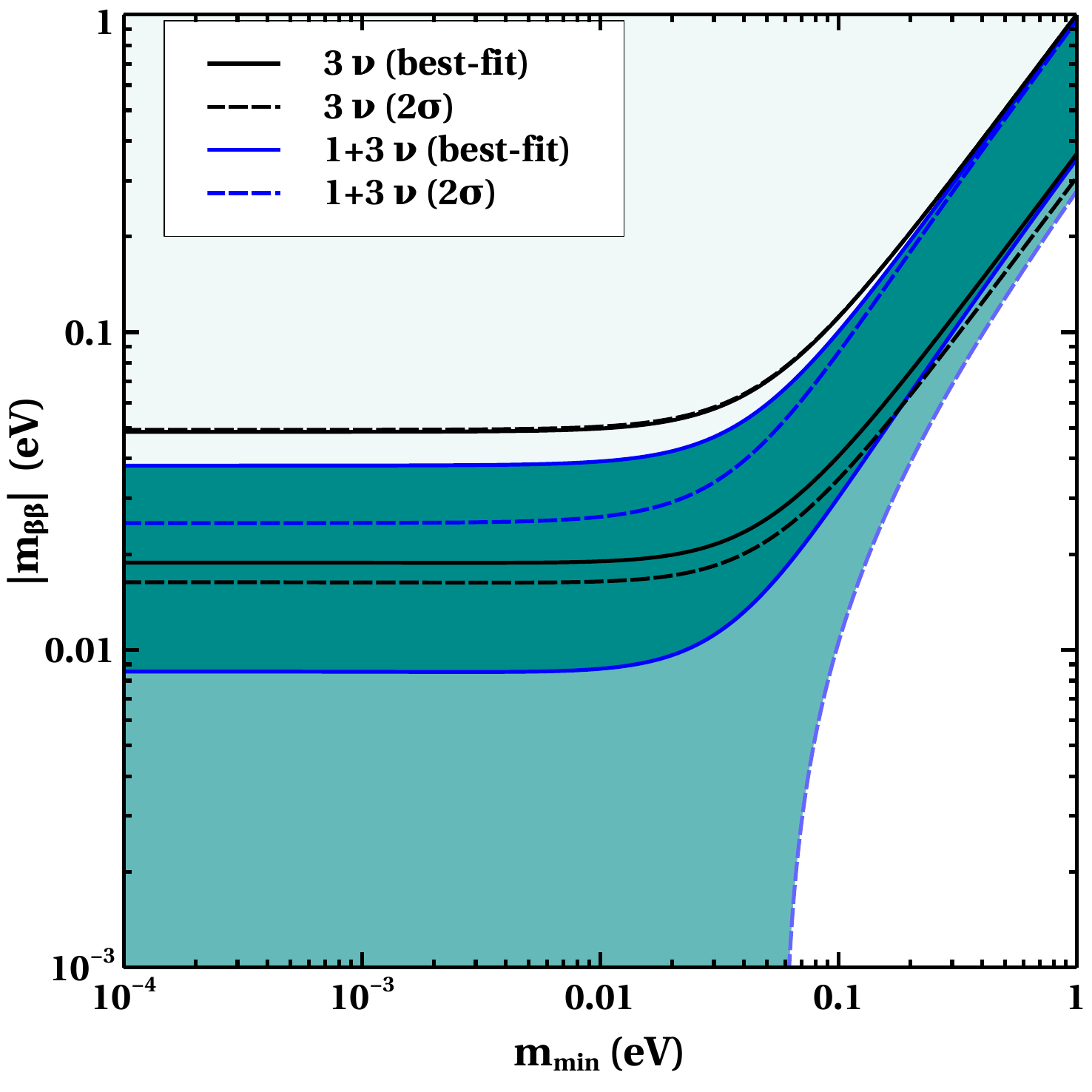}
\includegraphics[scale=.5]{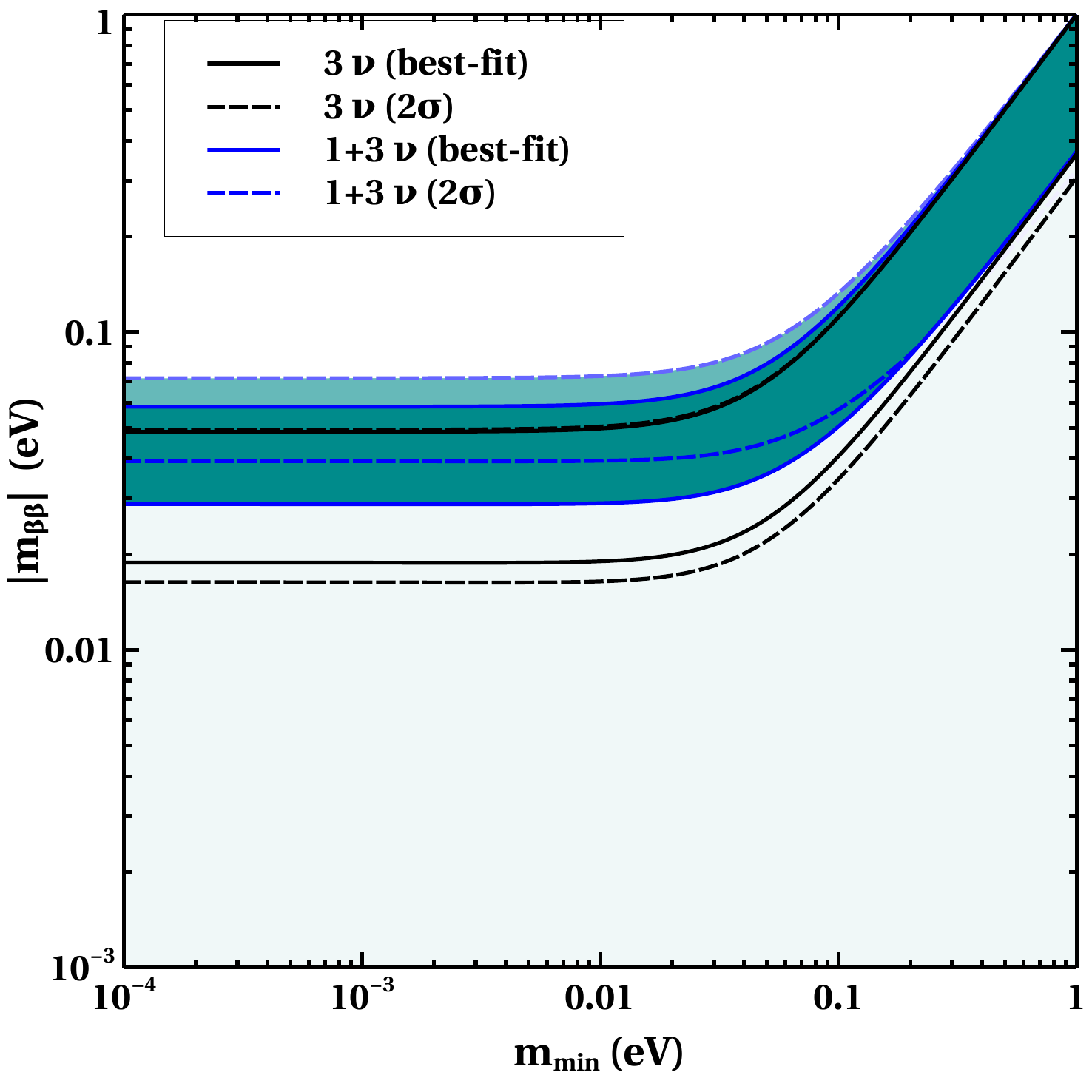}
\caption{ The allowed ranges in the $|m_{\beta\beta}|$ - $m_{low}$ parameter space, both in the standard three neutrino picture and with one sterile neutrino for the 1+3 IO 
case. Top left panel upper and lower band range are obtained in the mass term combination of (+,+,-) and (-,-,+), while right panel upper and lower band range are obtained in 
the mass term combination of (+,+,-) and (-,-,-). Bottom upper and lower band range are obtained in the mass term combination of (+,+,+) and (-,-,+). The signs indicate 
the signs of $e^{i\alpha_{21}}, e^{i\alpha_{31'}}, e^{i\alpha_{41}} = \pm$ 1, these are the extreme cases which determine the minimum and maximum value of 
$|m_{\beta\beta}|$ in which CP is conserved}
\end{figure} 

\end{document}